\documentstyle[aps,epsf,preprint]{revtex}
\newcommand{\be}{\begin{equation}}
\newcommand{\ee}{\end{equation}}

\begin{document}
\preprint{KAIST-TH 2001/04}
\draft

\title{Muon anomalous magnetic moment, $B\rightarrow X_s \gamma$
and  dark matter detection in the string models with dilaton domination
}
\author{$^a$ S. Baek ,~~ $^b$ P. Ko , ~~$^b$ Hong Seok Lee}

\address{
$^a$ Department of Physics, National Taiwan University \\
Taipei 10764,  Taiwan
\\
$^b$ Department of Physics, KAIST \\
 Taejon 305-701, Korea \\
}

\maketitle
\thispagestyle{empty}

\tighten

\begin{abstract}
We consider the muon anomalous magnetic moment $a_{\mu}$
in the string models with dilaton domination with two different
string scales : the usual GUT scale and the intermediate scale.
After imposing the direct search limits on the lightest neutral
Higgs and SUSY particle masses and the lightest neutralino LSP,
the $a_{\mu}^{\rm SUSY}$ is predicted to be less than $ 65~ (55)
\times 10^{-10}$ for $M_{string} = 2 \times 10^{16}$ GeV ($3
\times 10^{11}$ GeV). If we further impose the $B\rightarrow X_s
\gamma$ branching ratio, the predicted $a_{\mu}^{\rm SUSY}$ becomes lower to  
$35 \times 10^{-10}$ for intermediate string scale. 
The resulting LSP - proton scattering cross section is less than
$\sim 10^{-7}$ pb, which is below the sensitivity of the current direct 
dark matter search experiments, but could be covered by future experiments. 
\end{abstract}



\newpage

\narrowtext 
\tighten

The anomalous magnetic dipole moment (MDM) of a muon,
$a_{\mu} \equiv ( g_{\mu} - 2 )/2$, is
one of the best measured quantities with clean theoretical understanding.
Recently, the Brookhaven E821 announced a new data on
$a_{\mu}$ \cite{Brown:2001mg}:
\begin{equation}
  \label{eq:amuexp}
  a_{\mu}^{\rm exp} = (11659202 \pm 14 \pm 6) \times 10^{-10}.
\end{equation}
On the other hand, the SM prediction for this quantity has been calculated
through five loops in QED and two loops in the electroweak interactions
\cite{Czarnecki:2001pv} :
\begin{equation}
  \label{eq:amusm}
  a_{\mu}^{\rm SM} = ( 11659159.7 \pm 6.7 ) \times 10^{-10}.
\end{equation}
This new BNL result is $2.6 \sigma$ larger than the SM prediction,
and the difference between the two,
\begin{equation}
  \label{eq:amunew}
\delta a_{\mu} \equiv
a_{\mu}^{\rm exp} - a_{\mu}^{\rm SM} = (43 \pm 16) \times 10^{-10},
\end{equation}
could be a signal of new physics beyond the standard model (SM),
although the statistical significance has to be improved further and
hadronic uncertainties in the vacuum polarization and light - light
scattering in the SM prediction should be examined more carefully
\cite{Yndurain:2001qw}.
There have been many discussions on $\delta a_{\mu}$ in the context of
supersymmetric models \cite{susy} \cite{susynew}, in non-supersymmetric models
\cite{nonsusy}, as well as in a model independent way \cite{Einhorn:2001mf}.

In this letter, we study the muon anomalous magnetic moment and
$B\rightarrow X_s \gamma$ in the string models with dilaton
domination scenario. If we imposed only the direct search limits
for Higgs and SUSY particles and $B\rightarrow X_s \gamma$, there
remains a large parameter space in the $( m_{3/2}, \tan\beta )$.
But in most of this parameter region, we find that the $\delta
a_{\mu} \equiv a_{\mu}^{\rm SUSY}$ turns out to be small compared
to the BNL E821 result. There remains only a limited region that
is consistent with both $a_{\mu}$ and other direct and indirect
constraints on $( m_{3/2}, \tan\beta )$.  We also study the
neutralino LSP ($\chi_1^0$) -- proton cross section
$\sigma_{\chi^0_1 p}$, which is relevant to the direct search for
neutralino dark matter search experiments.

Let us briefly discuss the string models with dilaton domination
scenario. In the present, the most popular extension of the
standard model (SM) is the minimal supersymmetric standard model
(MSSM), which is presumably a low energy effective theory of more
fundamental theory such as superstring or $M$ theory. In weakly
interacting perturbative string models, the SUSY breaking can be
parameterized in terms of auxiliary fields of two different kinds
of quantities : dilaton superfield $S$ and moduli superfields
$T_i$ \cite{Brignole:1994dj}. In principle, the $F$ terms of all
these fields could contribute to SUSY breaking, and the soft terms
of the low energy theory will depend on $\langle F_S \rangle$ and
$\langle F_{T_i} \rangle$.
In this work, we will concentrated on the dilaton domination
scenario where only $\langle F_S \rangle$ plays an important role 
in SUSY breaking for the following reasons.

The dilaton domination scenario is very intriguing in
phenomenological senses, since it provides a solid ground for the
universality of sfermion masses at the string scale, thereby
solving the SUSY flavor problems. On the contrary, the so-called
minimal supergravity scenario, although this model is a kind of
benchmark in the SUSY phenomenology, has no such universality when
one goes beyond the minimal K\"{a}hler metric. For example, the
quantum supergravity corrections can generate a significant
non-universality in the sfermion mass terms \cite{Choi:1998de}.
However, there is an unsatisfactory aspect of dilaton domination
scenario with $M_{\rm string} \sim 2 \times 10^{16}$ GeV : the
whole parameter space is excluded by charge and color breaking
(CCB) minima \cite{Casas:1996wj}. Although one could assume that
our universe lives in the metastable state for long time before it
decays into a true minima breaking charge and color, it would be
nice if this problem could be solved within the particle physics
context. The authors of Ref.~\cite{Abel:2000bj} showed that this
is in fact achieved in intermediate scale string models with
dilaton domination scenario. When one lowers the string scale to
the intermediate scale, the CCB constraints becomes considerably
weaker, and the resulting string models have an ample parameter
space  consistent with the phenomenological constraints from
unobserved Higgs and SUSY particles. Thus the intermediate scale
string models with dilaton domination scenario were advocated as a
phenomenologically attractive scenario providing a natural
solution to the SUSY flavor problem within the string theory
context.

For long time, the fundamental scale of the string theory was
thought to be close to the Planck scale so that their low energy
implications were doomed to be irrelevant, except that some string
moduli can play an important role in cosmology. However this
picture has drastically changed after the so-called second string
revolution \cite{Ibanez:2000bn}.
The string scale is now thought to be anywhere from the Planck
scale (which is too high from the particle physics point of view)
down to the electroweak scale (which is a good news for particle
physics experiments) \cite{scale}. The important ingredient for
this is the existence of solitonic objects called $D$ branes on
which open string ends can attach \cite{Polchinski:1995mt}. Then
SM fields are the excitation modes of open string that can be
confined to the $D 3$ branes. On the other hand, the gravity is
the zero mode of a closed string so that it can propagate in the
bulk. This can make the fundamental scale different from the
Planck scale. Also the presence of $D$ branes reduce the number of
SUSY generators and it helps us construct realistic (MS)SM like
4-dimensional particle physics models.

Recently a class of Type-I string models were constructed by
orientifolding Type IIB string models \cite{Abel:2000bj}. These
new classes of Type-I string models differ from the old weakly
coupled string models in two important aspects. First of all, the
string scale can be arbitrary in principle, and one can make some
physical arguments for choosing a particular string scale. In this
context, the authors of Ref.~\cite{Abel:2000bj} argued that the
intermediate string scale is natural in many senses : hidden
sector and gravity mediated SUSY breaking scenarios, strong CP
problem, neutrino masses in the see-saw mechanism, and gauge
coupling unification, etc.. Secondly there appear one more moduli
field,  the Ramond-Ramond superfields $M_i$ associated with the
blowing up of orbifold singularities in orientifold constructions.
This new object appears in the Type-I string models with D branes,
and is important in $U(1)$ anomaly cancellation and generation of
the FI terms as well as string axions as a solution to the strong
CP problem \cite{Abel:2000bj}. In principle, the $F$ terms of all
the fields $S, T_i$ and $M_i$ could contribute to SUSY breaking,
and the soft terms of the low energy theory will depend on
$\langle F_S \rangle$, $\langle F_{T_i} \rangle$ and $\langle
F_{M_i} \rangle$. The generic forms of the soft terms in Type-I
string models were described by Allanach et al.
\cite{Abel:2000bj}. in Type I string models so that one has to
include the anomaly mediation and the loop effects in the
K\"{a}hler potential. However, these loop effects are not well
known yet. Therefore they ignored the effects of nonvanishing
$\langle F_{T_i} \rangle$ and $\langle F_{M_i} \rangle$, and
concentrated on the dilaton domination scenario where only
$\langle F_S \rangle$ plays an important role in SUSY breaking. In
this work, we consider only the dilaton domination limit for two
different string scales, $M_{string} = 2 \times 10^{16}$ GeV and
the intermediate string scale $\sim 10^{11}$ GeV.

Some phenomenological aspects of this class of models with
intermediate string scale have begun to be explored. The gauge and
Yukawa coupling unifications and Higgs and SUSY particle spectra
were discussed in Ref.~\cite{Abel:2000bj}. Also it was pointed out
that the initial scale has a very interesting implication for dark
matter search experiments \cite{Gabrielli:2001uy} \cite{cerdano}. 
The authors of Ref.~\cite{Gabrielli:2001uy} used the minimal 
supergravity type boundary conditions for the soft SUSY breaking 
terms at the intermediate scale :
\[
m_0,~~~M_{1/2} = - A ,
\]
where $m_0$ and $M_{1/2}$ are the universal scalar and the gaugino
mass parameters, and $A$ is the universal trilinear couplings.
This model is not the same as the model we consider in this work
(see Eq. ~(4).) If one starts the RG running from the intermediate
scale with the above boundary conditions for the soft SUSY
breaking parameters, the size of the $\mu$ parameter becomes lower
and the Higgsino component of the lightest neutralino LSP may
increase, depending on the choice of $M_0$ and $M_{1/2}$. This
could enhance the neutralino LSP couplings to Higgs bosons so that
the spin independent neutralino - nucleus scattering cross section
much larger than the binolike LSP case.

With such phenomenologically interesting aspects 
as well as theoretical motivations for 
string models with dilaton domination scenario, it would be worthwhile to
study phenomenological aspects of these models in more detail.
Now the new BNL data on the $a_{\mu}$
began to probe the electroweak and SUSY loop effects on $a_{\mu}$.
Also it is well known that $B\rightarrow X_s \gamma$ branching ratio puts
strong constraint on SUSY models, but this constraint was not considered
explicitly in the context of the intermediate scale string models with
dilaton dominance scenario. Therefore,  we study these two observables
within the intermediate scale string models with dilaton domination scenario,
in addition to the direct search limits considered in the previous study
\cite{Abel:2000bj}.

Assuming that the cosmological constant vanishes 
\footnote{The recent determination of the cosmological parameters strongly
  favors the presence of a sizeable dark energy ($\Omega_{\Lambda}$) in the
  Universe. One can apply the formulae in Refs.~\cite{Brignole:1994dj} 
  in order to 
  get soft parameters when the nonvanishing cosmological constant. 
  However its size is too small in the TeV scale region we are 
  considering so that its presence is practically unimportant. 
   } 
and $R$ parity is 
conserved, the soft terms in the dilaton domination scenario are given by
\cite{Abel:2000bj}
\begin{equation}
  M_{1/2}   =  \sqrt{3} \, m_{3/2} = - A.
\end{equation}
Here, $m_{3/2}$ is the gravitino mass parameter which is equal to the
universal scalar mass $m_0$, and $\tan\beta$ is another free parameter of this 
model. We have ignored the gauge group dependent loop correction effects 
in the gaugino mass parameter $M_a$.  Therefore, the soft terms of Type I
string models in the dilaton domination scenario are identical to the weakly
coupled heterotic string models in the dilaton domination scenario. The only
difference of these two models are the scale at which the RG running starts,
and this effect was shown to be very important \cite{Abel:2000bj}
\cite{Gabrielli:2001uy}.

We vary two input parameters $m_{3/2}$ upto 400 GeV and $\tan\beta$ upto 50, 
and do the standard renormalization group analysis with the above
boundary conditions at some string scales $M_{\rm string}$. Then
the particle spectra and mixings are determined with resulting
parameters at the electroweak scale. For the string scale $M_{\rm
string}$ and gauge coupling unification, we consider the following
three possibilities \footnote{We used one loop RG equations for
the runnings for gauge and Yukawa couplings so that the gauge
coupling unifies at a slightly higher scale than the scale
obtained in Ref.~\cite{Abel:2000bj}.} :
\begin{itemize}
  \item P1 : the string scale at $M_{\rm string} = 2 \times 10^{16}$ GeV
  with gauge coupling unification
  \item P2 : the string scale at $M_{\rm string} = 3 \times 10^{11}$ GeV
  without gauge coupling unification
  \item P3 : the string scale at $M_{\rm string} = 3 \times 10^{11}$ GeV
  with gauge coupling unification by adding extra leptons :
  $\left( 3 \times E_R + 2 \times L \right)$ and their vectorlike partners
\end{itemize}
The motivation for this  intermediate scale is given by Abel 
{\it et al.} in Ref.~\cite{Abel:2000bj} : 
hidden sector and gravity mediated SUSY breaking scenarios, 
strong CP problem, neutrino masses in the see-saw mechanism, and gauge 
coupling unification at this scale by adding vectorlike leptons, etc..
As mentioned before, the whole parameter space of the first case P1 is
excluded by CCB constraints  \cite{Casas:1996wj}, but not in the cases of 
intermediate scale, P2 and P3 \cite{Abel:2000bj}. Also, in the case P2, it 
was shown that the charged LSP constraint imposes $\tan\beta < 28~$ so that
the bottom-tau Yukawa unification is not possible \cite{Abel:2000bj}.
On the other hand, if one adds extra vectorlike leptons in order to achieve
intermediate scale gauge coupling unification, the charged LSP constraint
becomes much milder and the bottom-tau Yukawa unification becomes possible
for any values of $\tan\beta$. We also assume radiative electroweak symmetry
breaking condition. 

We impose the direct search limits on the lightest Higgs mass ~\cite{drees}
\begin{eqnarray*}
m_{h} > 93.5 +15x +54.3x^2 -48.4x^3-25.7x^4 +24.8x^5 -0.5 
~{\rm GeV}~,
\end{eqnarray*}
where $x=\sin^2(\beta-\alpha)$ and $\alpha$ is the mixing angle of the
CP-even Higgs bosons, as well as on the SUSY particle masses :
\begin{eqnarray*}
 m_{\chi^{\pm}} > 84 ~{\rm GeV}~\cite{ch}~, \qquad
&m_{\chi^0_1} > 31 ~{\rm GeV}~\cite{ch}~,  \qquad
&m_{\tilde{g}} > 300 ~{\rm GeV}~\cite{glue}~,  \\
m_{\tilde{t_1}} >  83~{\rm GeV}~\cite{st1}~, \qquad 
&m_{\tilde{\tau_1}} > 72~{\rm GeV}~\cite{stau1}~. \qquad
&
\end{eqnarray*}
It turns out that the lightest neutral Higgs and the lighter stau mass
limits are most severe constraints compared to others.
In the most parameter space of our model, we have the decoupling case 
$m_A^2 \gg m_Z^2$. Therefore, $\sin^2(\beta-\alpha) \simeq 1 $ and the 
interaction of the lightest CP even Higgs boson mass is almost SM-like, 
so $m_h > 113.5$~GeV 
is a pretty good approximation in the most parameter space region.
We also impose the indirect constraints from $a_{\mu}$~\cite{Brown:2001mg}
and $B\rightarrow X_s \gamma$~\cite{Cleo}
at the $2 \sigma$ level :
\begin{eqnarray}
11 \times 10^{-10} & < &  a_{\mu}^{\rm SUSY}  < 75 \times 10^{-10} \\
2.18 \times 10^{-4} & < & B(B \rightarrow X_s \gamma ) < 4.10 \times
10^{-4}.
\end{eqnarray}
We also excluded the region where the LSP is charged.
If we relax the assumption that the $R$ parity is conserved, the charged LSP
region may not be excluded and the allowed parameter region would be wider.
However, in such a case, there would be additional contributions to the
$a_\mu^{\rm SUSY}$ as well as to the 
$B\rightarrow X_s \gamma$ branching ratio from
$R$ parity violating interactions, which would make the whole analysis quite
complicated. With $R$ parity conservation, the LSP will be a good candidate
for the cold dark matter, the detection of which are actively pursued now
at many places.
We will study the neutralino LSP - proton scattering cross section in the
allowed parameter space in the string models with dilaton domination with
different string scales.

It is well known that the sign of $a_{\mu}^{\rm SUSY}$ 
is correlated with the sign of
$\mu$, and Eq.~(\ref{eq:amunew}) implies that $\mu > 0$ and relatively large
$\tan\beta$ is preferred and SUSY particles cannot be too heavy.
On the other hand, the SM explains the $B \rightarrow X_s \gamma$ rate very
well so that there cannot be significant new physics contribution to it, if
the new physics contribution has the same sign as the SM amplitude. This
means that the chargino -- stop contributions to $B \rightarrow X_s \gamma$
should interfere destructively with the SM and the charged Higgs contributions
in order to satisfy the $B\rightarrow X_s \gamma$ constraint \cite{bsg},
which would lead to interesting consequences in other $B$ decays.
In this work, we used the NLO calculations for SM contributions and the LO
results for SUSY contributions to $B \rightarrow X_s \gamma$. More complete
analysis including the SUSY NLO effects enhanced by large $\tan\beta$ will
be discussed in a separate publication \cite{future}.  And some
care should be exercised when we consider the $B \rightarrow X_s \gamma$
constraint in SUSY models with large $\tan\beta$ region, which is relevant to
the BNL data on $a_{\mu}^{\rm SUSY}$. 
If $\tan\beta$ is large, then SUSY QCD corrections
to the bottom Yukawa couplings can be $\sim O(1)$, and one has to make
resummation of such enhanced contributions to $B \rightarrow X_s \gamma$.
Such attempts were made recently and it was found that the SUSY contribution
to $B \rightarrow X_s \gamma$ for $\mu > 0 $ (which is selected by the
$a_{\mu}^{\rm SUSY}$)  can be enhanced by more than $\sim 50 \%$ for large
$\tan\beta \sim 30$ \cite{largetan}.  Therefore some points below the
$B \rightarrow X_s \gamma$ lower bound may be within the bound after large
$\tan\beta$ terms are appropriately resummed.

In Figs.~1 and 2, we show the allowed regions in the $(m_{3/2}, \tan\beta)$
plane with $\mu > 0$ for the case P1 and the correlation between the
$a_{\mu}^{\rm SUSY}$ and $B\rightarrow X_s \gamma$  branching ratio therein,
respectively. In this case, the whole region is not compatible with the
absence of CCB minima condition, which we ignore for the moment.
In Fig.~1, the shaded and the dark regions are excluded by the charged
LSP and direct search limits on SUSY and Higgs particles, respectively.
It turns out that the lower bounds on the lightest neutral Higgs and the
lighter stau masses are the most stringent one.
The remaining parameter space is consistent with
$B\rightarrow X_s \gamma$ branching ratio which is shown by slanted lines.  
Also the constant $a_{\mu}^{\rm SUSY}$ contours for $a_{\mu}^{\rm SUSY} =
(11,27,43) \times 10^{-10}$ are shown in the same parameter space.



From Fig.~2, we observe that the $a_{\mu}^{\rm SUSY}$ 
and $B\rightarrow X_s \gamma$
branching ratio is anticorrelated with each other. However, these two are not
inversely proportional to each other. Rather the actual correlation is
approximately a parabola. The reason is the following. For large $\tan\beta$,
\begin{eqnarray*}
a_{\mu}^{\rm SUSY} & \sim & \mu \tan\beta \\
{\cal M}( b \rightarrow s \gamma ) & \sim & ({\rm SM ~Amp.}) +
( \# ) \times \mu \tan\beta,
\end{eqnarray*}
where $\#$ is a number depending on SUSY parameters in the loop integral.
Since the branching ratio for $B\rightarrow X_s \gamma$  is obtained by
squaring the amplitude, we will have quadratic dependence of the
$B\rightarrow X_s \gamma$ branching ratio on $a_{\mu}^{\rm SUSY}$.
Let us note that if we choose $\mu > 0$ in order to explain the BNL data,
then $B\rightarrow X_s \gamma$ branching ratio turns out to be in the 
relatively lower side. 
A larger $a_{\mu}^{\rm SUSY}$ would be 
eventually constrained by the direct search limits on
Higgs and SUSY particles (especially the lighter stau) and the
$B\rightarrow X_s \gamma$ branching ratio.

Similar plots for the cases P2 and P3 (the intermediate string scale without
and with gauge coupling unification) are shown in Figs.~3, 4 and Figs.~5, 6,
respectively. In the P2 case, the allowed region becomes significantly reduced
compared to the cases P1 or P3
mainly because of the charged LSP constraint.
The resulting $a_{\mu}^{\rm SUSY}$ becomes somewhat smaller compared to
the case P1, partly because $\mu$ parameter gets smaller in the intermediate
scale string models, but mainly because the allowed region is too small.
$a_{\mu}^{\rm SUSY}$ cannot be larger than 
$30 \times 10^{-10}$ in the P2 case.
For the case P3 (Figs.~5 and 6), the allowed parameter space becomes
significantly larger compared to the case P2. 
The $a_{\mu}^{\rm SUSY}$ can be as large as $55 \times 10^{-10}$,
if we ignore $B\rightarrow X_s \gamma$ branching ratio. 
However, if we impose the $B\rightarrow X_s \gamma$ constraint, 
the possible $a_{\mu}^{\rm SUSY}$ cannot be larger than $35 \times 10^{-11}$.
In any case, the larger $a_{\mu}$ tends to prefer the smaller branching 
ratio for $B\rightarrow X_s \gamma$. 

Let us consider the effect of the muon anomalous magnetic moment on the
neutralino LSP - proton scattering cross section $\sigma_{\chi^0_1 p}$
~\cite{dm},
which is relevant to the dark matter search experiments.
For a given string scale, the universal scalar mass $m_0(=m_{3/2})$ at the string
scale is approximately proportional to $\tan\beta$, when we impose the
$a_{\mu}^{\rm SUSY}$ and $B\rightarrow X_s \gamma$ constraint. 
Therefore, the SUSY
masses tend to increase as $\tan\beta$ increases for a fixed $M_{\rm string}$.
On the other hand, for a fixed $\tan\beta$, the smaller $m_{3/2}$ is favored
by $a_{\mu}$ data, the SUSY particle masses becomes lighter and the cross
section will increase.
These behaviors can be seen from Figs.~7, 8 for the case P1.
In Fig.~7, we show the constant contours for the cross section
$\sigma_{\chi^0_1 p}$ (in unit of pb) in the $(m_{3/2}, \tan\beta
)$ plane. The dashed curve in the region allowed by the
$B\rightarrow X_s \gamma$ constraint represents the $2 \sigma$ lower
bound to $a_{\mu}^{\rm SUSY}$, and the upper left part of this dashed 
curve is consistent with the new BNL data on $a_{\mu}$. Note that the
neutralino LSP - proton scattering cross section is  less than
the sensitivity of current direct dark matter search experiment (DAMA, CDMS)
($\sim 10^{-6}$ pb)~\cite{dmexp}, mainly because of the  
direct search limits on Higgs and SUSY particles as well as the
$B\rightarrow X_s \gamma$ branching ratio. Therefore the dark matter
search experiment cannot be complementary to the indirect constraint from
$B\rightarrow X_s \gamma$ in the string models with dilaton domination
with $M_{string} = 2 \times 10^{16}$ GeV.

In Fig.~8, we show the dependence of the cross section on
$\tan\beta$, along with the constant contours for $a_{\mu}^{\rm SUSY}$ in
unit of $10^{-10}$. The slanted lines represent the points
consistent with the $B\rightarrow X_s \gamma$ constraint. 
The left part is cut by the direct search limit and  
the right part of the slanted lines are cut by the charged LSP
constraints. The lower cut of the slanted region is due to the 
artificial cut at $m_{3/2} = 400$ GeV. 
As $a_{\mu}^{\rm SUSY}$ increases, the cross
section $\sigma_{\chi^0_1 p}$ also increases, since the scalar mass
parameter $m_{3/2}$ decreases. For fixed $a_{\mu}^{\rm SUSY}$, the
cross section is a decreasing function of $\tan\beta$, since
$m_{3/2}$ also increases when $\tan\beta$ increases and the
combined effects result in the decreasing cross section as a
function of $\tan\beta$. 

In Figs.~9 and 10, we show the same plots for the case P3.
(The case P2 is very similar to the case P3 for the neutralino LSP -
proton scattering cross section, except that the allowed parameter
region is small, and we do not show the plots separately.) In this
case, the cross section increases by a factor of $\sim 10$ compared
to the case P1. This is not because the Higgsino component of the LSP
increases like the model considered in Ref.~\cite{Gabrielli:2001uy}.
In our case, the gaugino and the scalar mass parameters are not
independent with each other, but they are tightly correlated via
Eq.~(4). In fact, the neutralino LSP in our model is still binolike,
not Higgsino like (see Fig.~11), even if the $\mu$ parameter decreases
compared to the case P1.
The reason why the cross section increases in the case P3
compared to the case P1 is that the squark mass becomes lower in
the P3 case, for the same reason why the $\mu$ parameter becomes
lower in the intermediate scale string models.
Still, the predicted cross section is less than $\sim 10^{-7}$
pb, which is below the current direct dark matter search limit
over the allowed parameter space. 
The reason is that the neutralino in the intermediate scale string 
models with dilaton domination scenario is mainly binolike, not 
Higgsino like as in Ref.~\cite{Gabrielli:2001uy}. 
Note that the model considered in this
work Ref.~\cite{Gabrielli:2001uy} is not exactly the same as the model
we considered. Contrary to the claim made therein, the characters of the
neutralino LSP is quite sensitive to the parameters $m_0$ and $M_{1/2}$,
as our discussions demonstrate.



In conclusion, we showed that the parameter space of the
intermediate scale Type I string models with dilaton domination
scenario are strongly constrained by the recent measurement of the
muon anomalous MDM $a_{\mu}$, and also partly by $B\rightarrow X_s
\gamma$ branching ratio with some reservation for large
$\tan\beta$. If we impose constraints from direct search limits on
Higgs and SUSY particles and $B\rightarrow X_s \gamma$ only, there
still remains an ample parameter space in $( m_{3/2},\tan\beta )$
plane which indicates that the SUSY flavor problem is ameliorated
in this model. However a substantial part of this parameter space
is ruled out by the lower bound of the new data on the muon
anomalous magnetic moment. The resulting $a_{\mu}^{\rm SUSY}$ lies
in the relatively low side mainly because of the $B \rightarrow
X_s \gamma$ constraint.
In P3 case $a_{\mu}^{\rm SUSY}$ is somewhat smaller than P1 case,
because the resulting $\mu$ parameter becomes
smaller. The neutralino LSP -- proton scattering cross section is
also constrained by the BNL $a_{\mu}^{\rm SUSY}$ data. The parameter region
resulting in the small cross section is partly removed by the lower
bound on $a_{\mu}^{\rm SUSY}$. Still, the resulting region is too small to be
sensitive to current direct DM search experiments.  But in the near future, 
CDMS at Soudan~\cite{cdms} or CRESST~\cite{cresst} will be able to cover 
a part of this region with $\sigma_{\chi_1^0 p} \sim 10^{-8}$.

\acknowledgements
We are  grateful to K. Choi, M. Drees, P. Gondolo, T. Goto, Jihn E. Kim
and W.Y. Song for useful  communications.
This work was supported in part by BK21 program,
DFG-KOSEF exchange program and by KOSEF SRC program through CHEP at
Kyungpook National University.



\begin{figure}
\centerline{\epsfxsize=9cm \epsfbox{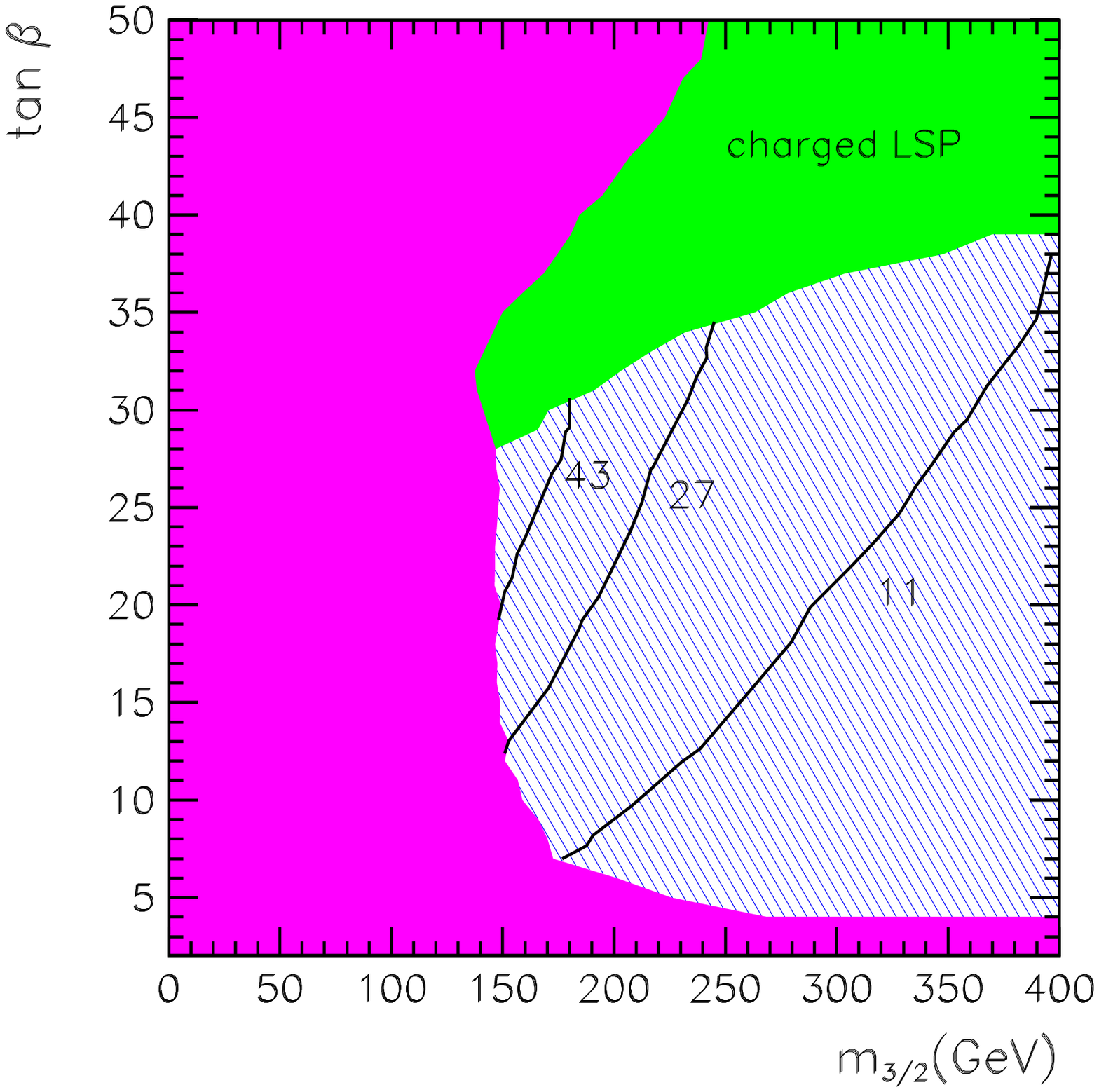}}
\caption{
The parameter space in the $(m_{3/2}, \tan\beta)$ space allowed by the
direct search limits and the charged LSP constraint for the case P1.
The region allowed by $B(B\to X_s \gamma)$ is denoted by slanted lines, and
the contours for constant $a_\mu^{\rm SUSY}$ (in unit of $10^{-10}$) are
shown in different curves. 
The dark region is excluded by direct search limit.
} \label{fig1}
\end{figure}

\begin{figure}
\centerline{\epsfxsize=9cm \epsfbox{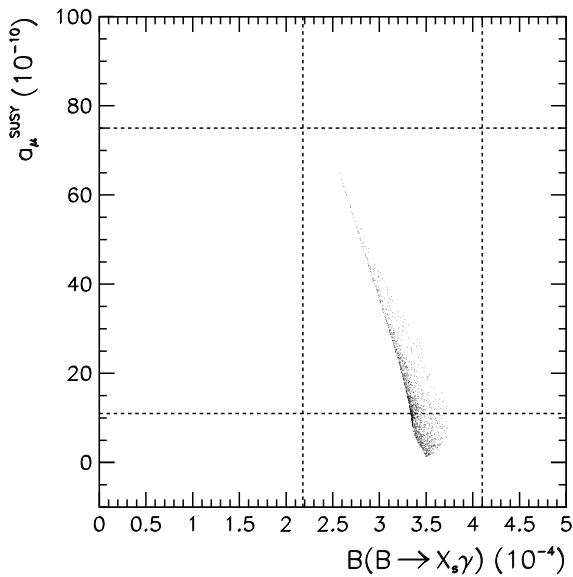}}
\caption{
The correlations between $a_{\mu}^{\rm SUSY}$ and $B\rightarrow X_s \gamma$
for the case P1. The vertical dashed lines represent the experimental data
for $B(B\to X_s \gamma)$, and the horizontal lines represent the bound of
$a_\mu^{\rm SUSY}$ to $2 \sigma$ level.
}
\label{fig2}
\end{figure}

\begin{figure}
\centerline{\epsfxsize=9cm \epsfbox{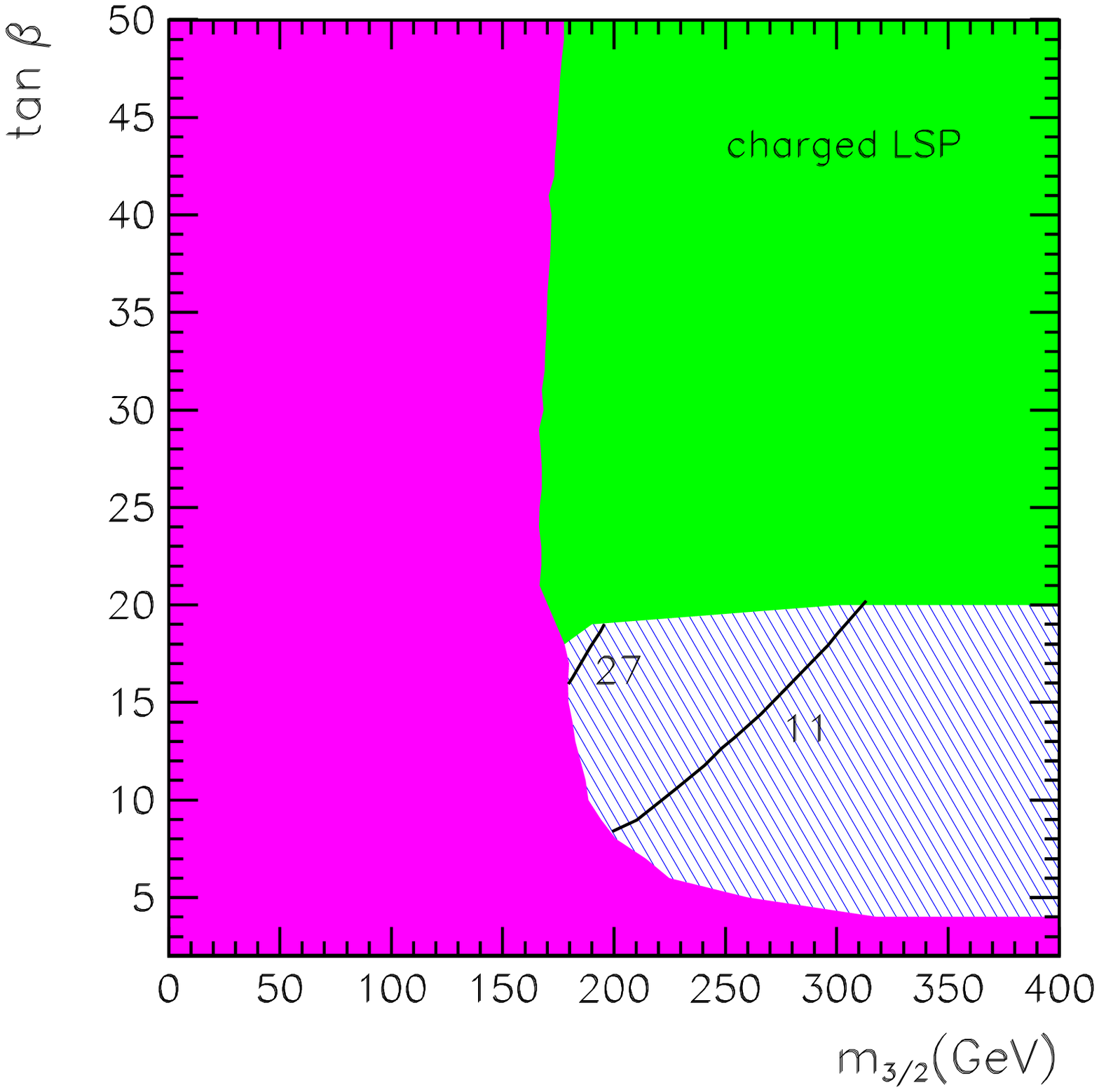}}
\caption{
The same plot as Fig.~1 for the case P2. The CCB minima constraint is
imposed,  although it is not shown explicitly.
}
\label{fig3}
\end{figure}

\begin{figure}
\centerline{\epsfxsize=9cm \epsfbox{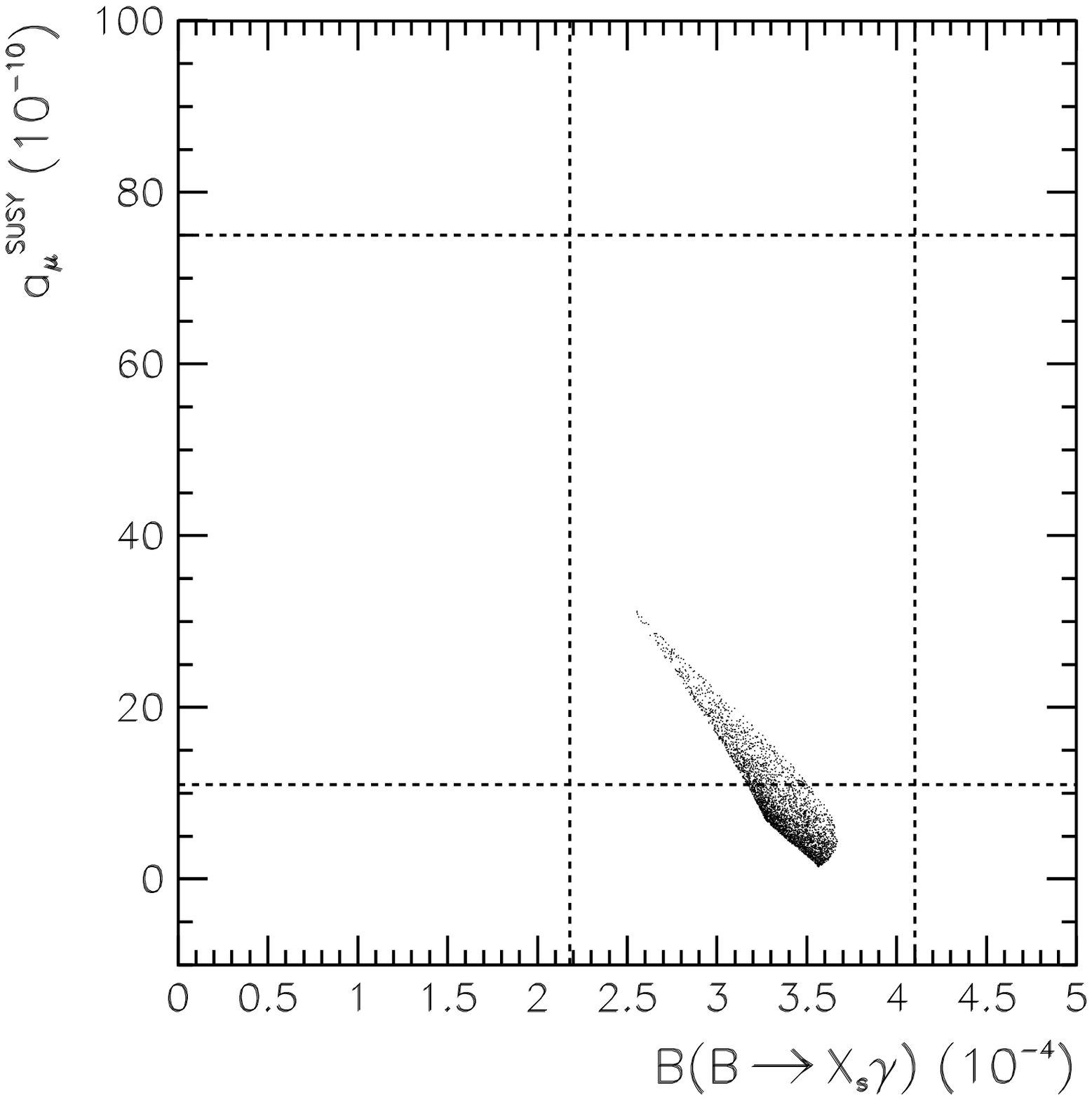}}
\caption{
The same plot as Fig.~2 for the case P2.
}
\label{fig4}
\end{figure}

\begin{figure}
\centerline{\epsfxsize=9cm \epsfbox{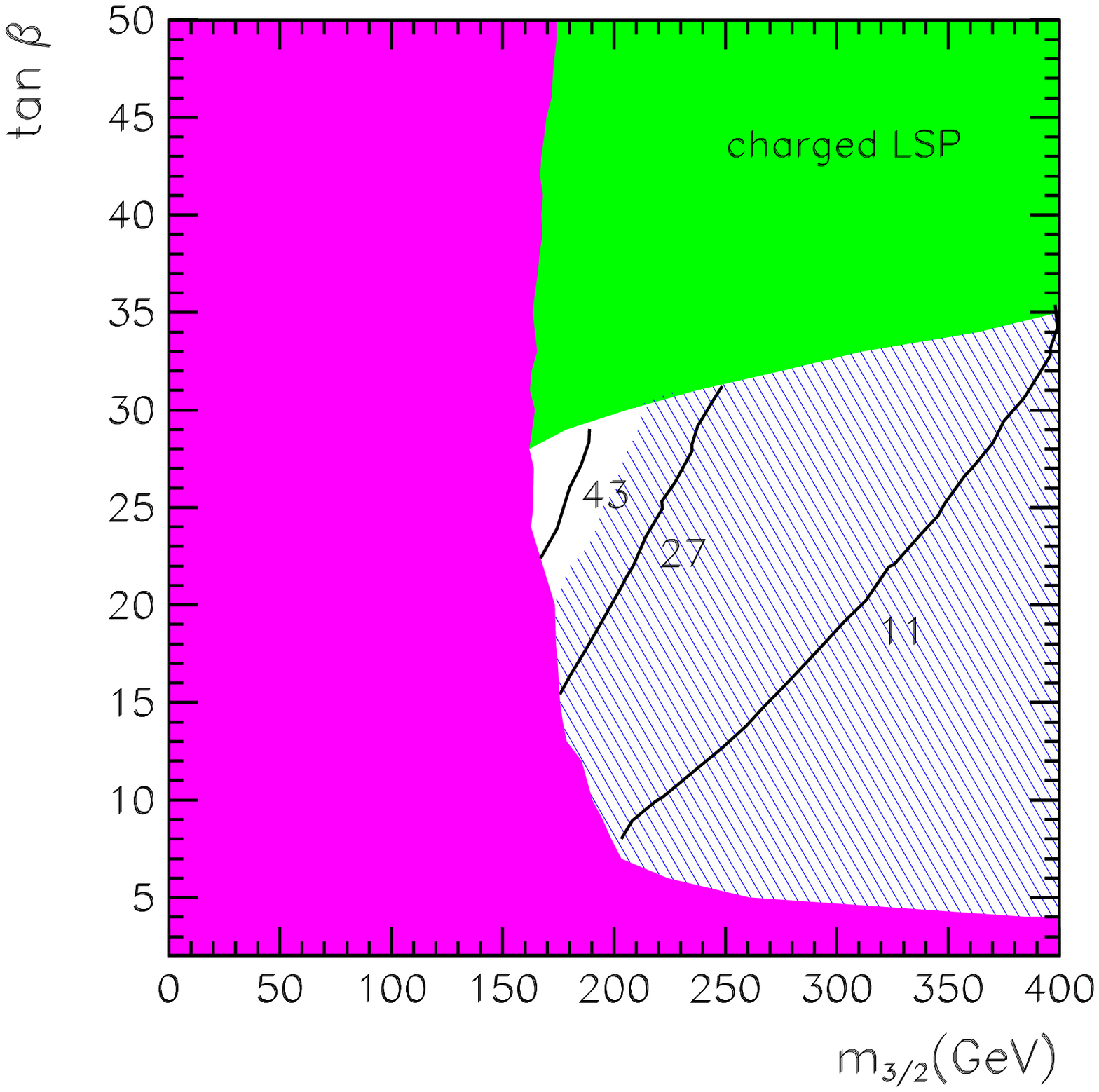}}
\caption{
The same plot as Fig.~1 for the case P3. The CCB minima constraint is
imposed,  although it is not shown explicitly.
}
\label{fig5}
\end{figure}

\begin{figure}
\centerline{\epsfxsize=9cm \epsfbox{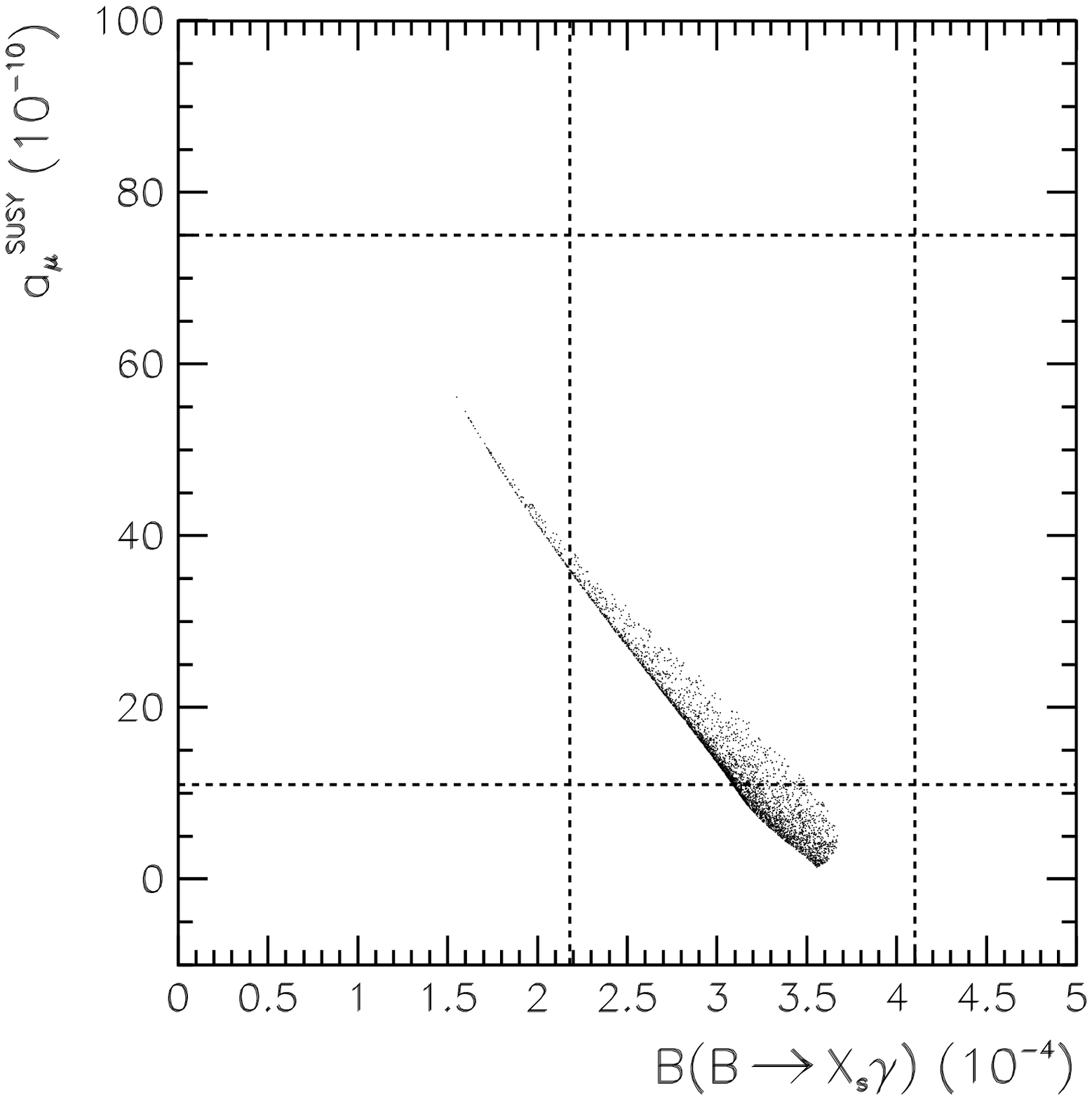}}
\caption{
The same plot as Fig.~2 for the case P3.
}
\label{fig6}
\end{figure}

\begin{figure}
\centerline{\epsfxsize=9cm \epsfbox{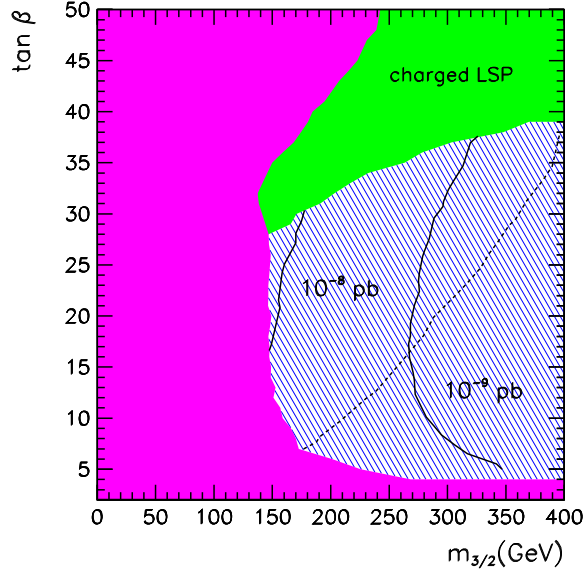}}
\caption{
The contours for the constant neutralino LSP -- proton scattering cross
section $\sigma_{\chi^0_1 p}$ (in pb) in the $(m_{3/2}, \tan\beta)$
plane for the case P1.
The dashed curve corresponds to the $2 \sigma$ lower bound on
$a_{\mu}^{\rm SUSY}$, the upper left part of which is consistent with
the BNL measurement of $a_{\mu}$.
} \label{fig7}
\end{figure}

\begin{figure}
\centerline{\epsfxsize=9cm \epsfbox{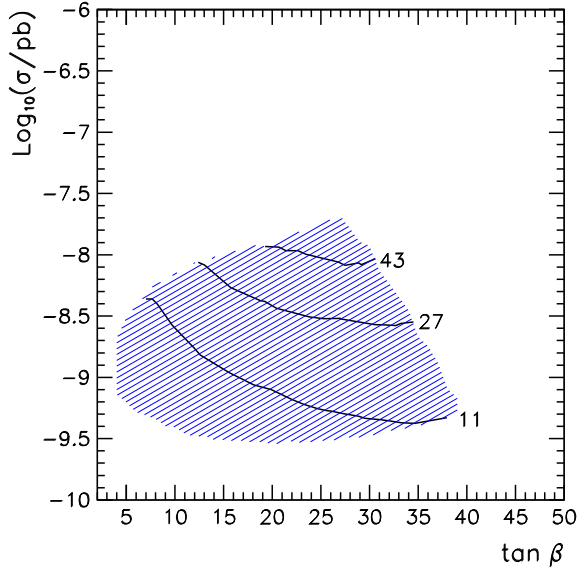}}
\caption{
The neutralino LSP -- proton scattering cross section $\sigma_{\chi^0_1 p}$
(in pb) dependence on $\tan\beta$ for the case P1.
The region allowed by $B(B\to X_s \gamma)$ is denoted by slanted lines, and
the contours for constant $a_\mu^{\rm SUSY}$ are shown in different curves.
The dashed horizontal line indicates the lowest sensitivity of DAMA
experiment.
} \label{fig8}
\end{figure}



\begin{figure}
\centerline{\epsfxsize=9cm \epsfbox{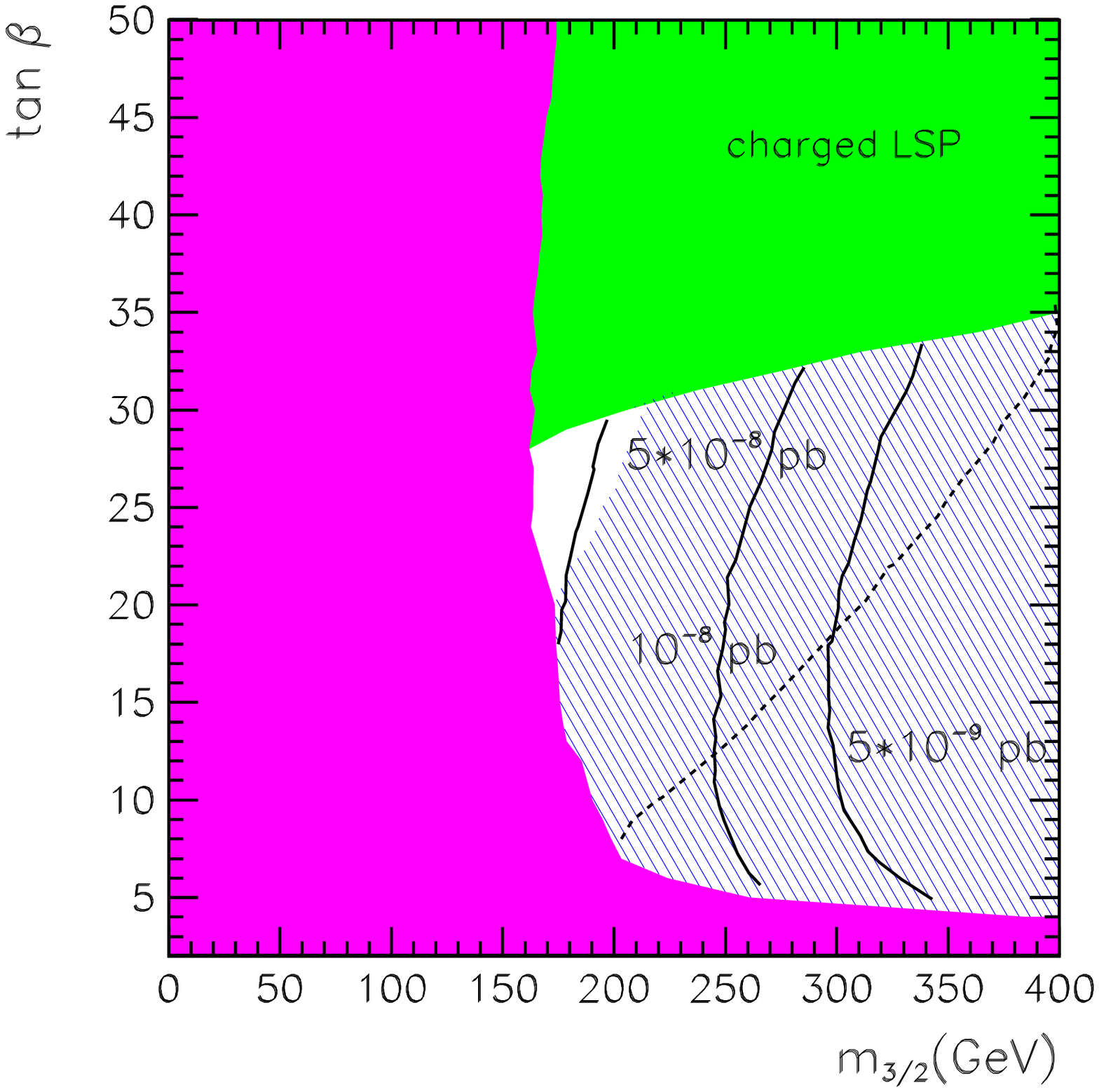}}
\caption{
The same plot as Fig.~5 for the case P3.
} \label{fig11}
\end{figure}

\begin{figure}
\centerline{\epsfxsize=9cm \epsfbox{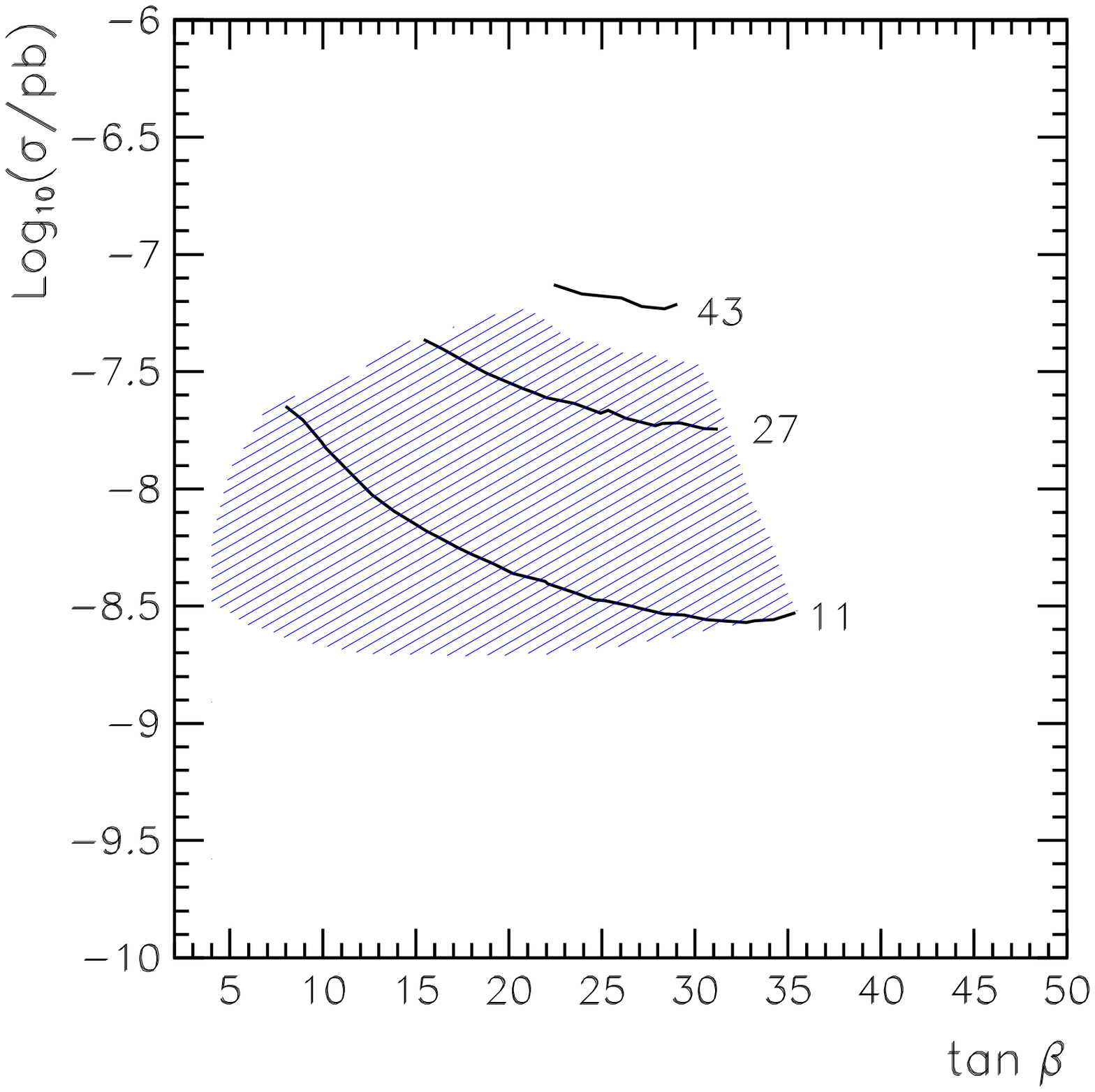}}
\caption{
The same plot as Fig.~6 for the case P3.
}
\label{fig12}
\end{figure}

\begin{figure}
\centerline{\epsfxsize=9cm \epsfbox{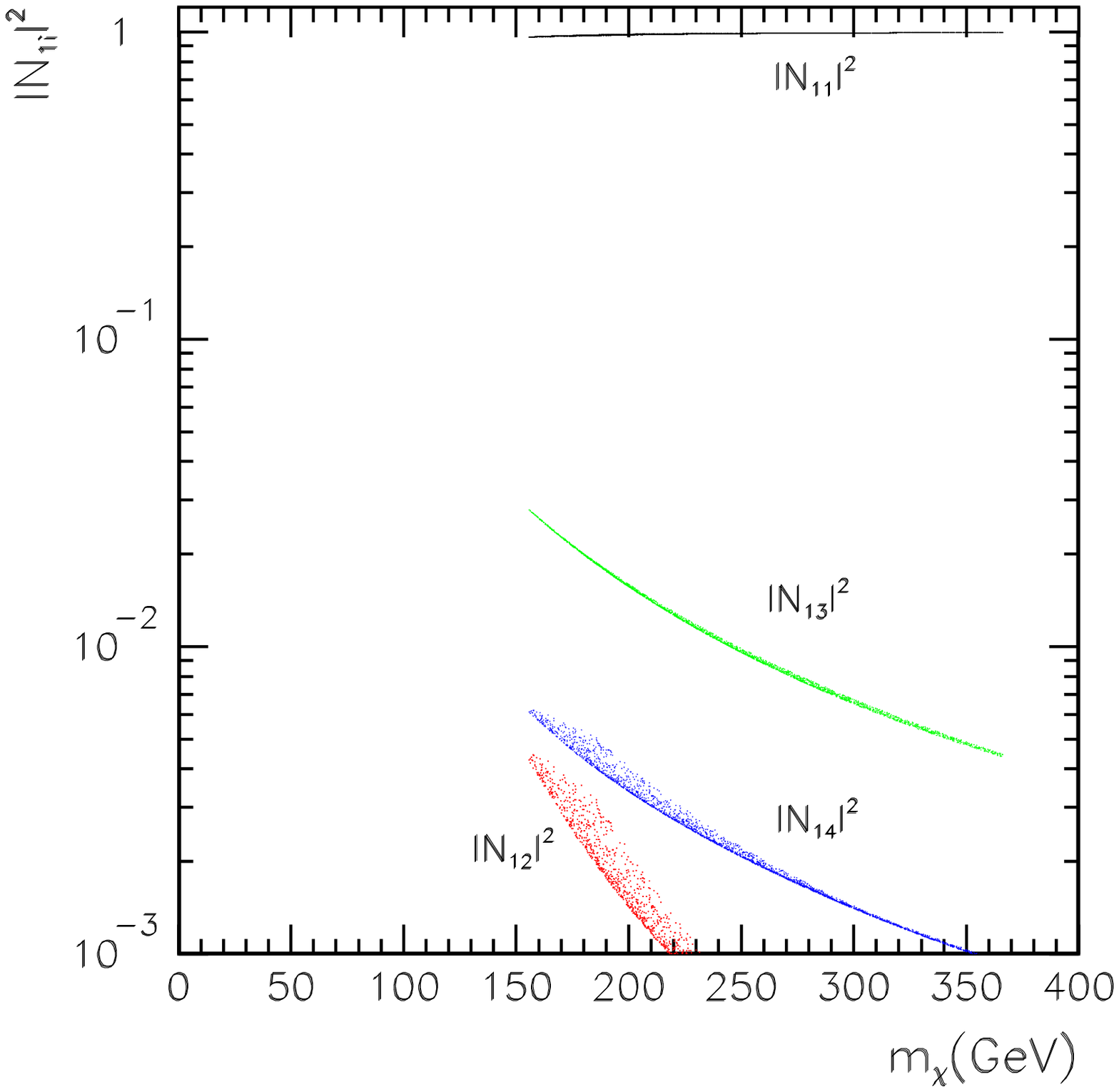}}
\caption{
The bino ($| N_{11} |^2$), wino ($| N_{12} |^2$ ), higgsino components
($| N_{13} |^2$ and $| N_{14} |^2$) of the lightest LSP as functions
of the LSP mass $m_{\chi}$ for the case P3.
}
\label{fig13}
\end{figure}


\vfil\eject


\begin{thebibliography}{99}

\bibitem{Brown:2001mg}
H.~N.~Brown {\it et al.}  [Muon $g-2$ Collaboration],
Phys. Rev. Lett. {\bf 86}, 2227 (2001).
\bibitem{Czarnecki:2001pv}
A.~Czarnecki and W.~J.~Marciano,
hep-ph/0102122, and references therein.
\bibitem{Yndurain:2001qw}
F.~J.~Yndurain,
hep-ph/0102312.
\bibitem{susy}
J.~L.~Lopez, D.~V.~Nanopoulos and X.~Wang,
   Phys.\ Rev.\ D {\bf 49}, 366 (1994) ;
U.~Chattopadhyay and P.~Nath,
   Phys.\ Rev.\ D {\bf 53}, 1648 (1996) ; 
T.~Moroi,
   Phys.\ Rev.\ D {\bf 53}, 6565 (1996) ;
M.~Carena, G.~F.~Giudice and C.~E.~Wagner,
   Phys.\ Lett.\ B {\bf 390}, 234 (1997). 
\bibitem{susynew}
L.~Everett, G.~L.~Kane, S.~Rigolin and L.~Wang,
   hep-ph/0102145 ;
J.~L.~Feng and K.~T.~Matchev,
   hep-ph/0102146 ;
E.~A.~Baltz and P.~Gondolo,
   hep-ph/0102147 ;
S.~Komine, T.~Moroi and M.~Yamaguchi,
   hep-ph/0102204 ;
K.~Choi, K.~Hwang, S.~K.~Kang, K.~Y.~Lee and W.~Y.~Song,
   hep-ph/0103048.
S.~P.~Martin and J.~D.~Wells,
   hep-ph/0103067 ;
S.~Komine, T.~Moroi and M.~Yamaguchi,
   hep-ph/0103182 ;
K.~Cheung, C.~Chou and O.~C.~Kong,
   hep-ph/0103183.
\bibitem{nonsusy}
S.~Davidson, D.~Bailey and B.~A.~Campbell,
   Z.\ Phys.\ C {\bf 61}, 613 (1994) ;
G.~Couture and H.~Konig,
   Phys.\ Rev.\ D {\bf 53}, 555 (1996) ;
P.~Nath and M.~Yamaguchi,
   Phys.\ Rev.\ D {\bf 60}, 116006 (1999) ;
M.~L.~Graesser,
   Phys.\ Rev.\ D {\bf 61}, 074019 (2000) ;
H.~Davoudiasl, J.~L.~Hewett and T.~G.~Rizzo,
   Phys.\ Lett.\ B {\bf 493}, 135 (2000) ;
D.~Chakraverty, D.~Choudhury and A.~Datta,
   hep-ph/0102180 ;
T.~Huang, Z.~H.~Lin, L.~Y.~Shan and X.~Zhang,
   hep-ph/0102193 ;
D.~Choudhury, B.~Mukhopadhyaya and S.~Rakshit,
   hep-ph/0102199 ;
S.~N.~Gninenko and N.~V.~Krasnikov,
   hep-ph/0102222 ;
K.~Cheung,
   hep-ph/0102238 ;
P.~Das, S.~K.~Rai and S.~Raychaudhuri,
   hep-ph/0102242 ;
T.~W.~Kephart and H.~Pas,
   hep-ph/0102243 ;
E.~Ma and M.~Raidal,
   hep-ph/0102255 ;
Z.~Xiong and J.~M.~Yang,
   hep-ph/0102259 ;
S.~K.~Kang and K.~Y.~Lee,
   hep-ph/0103064 ;
S.~Rajpoot,
   hep-ph/0103069 ;
S.~C.~Park and H.~S.~Song,
   hep-ph/0103072 ;
C.~Yue, Q.~Xu and G.~Liu,
   hep-ph/0103084 ;
R.~A.~Diaz, R.~Martinez and J.~A.~Rodriguez,
   hep-ph/0103050 ;
D.~A.~Dicus {\it et al.},
   hep-ph/0103126 ;
E.~O.~Iltan,
   hep-ph/0103105 ;
C.~S.~Kim, J.~D.~Kim and J.~Song,
   hep-ph/0103127.
\bibitem{Einhorn:2001mf}
M.~B.~Einhorn and J.~Wudka,
hep-ph/0103034.
\bibitem{Brignole:1994dj}
A.~Brignole, L.~E.~Ibanez and C.~Munoz,
Nucl.\ Phys.\ B {\bf 422}, 125 (1994) ;
Erratum-ibid.B {\bf 436}, 747 (1995) ;
A.~Brignole, L.~E.~Ibanez, C.~Munoz and C.~Scheich,
Z.\ Phys.\ C {\bf 74}, 157 (1997)
[hep-ph/9508258]
\bibitem{Choi:1998de}
K.~Choi, J.~S.~Lee and C.~Munoz,
Phys.\ Rev.\ Lett.\ {\bf 80}, 3686 (1998).
\bibitem{Casas:1996wj}
J.~A.~Casas, A.~Lleyda and C.~Munoz,
Phys.\ Lett.\ B {\bf 380}, 59 (1996) ;
J.~A.~Casas, A.~Lleyda and C.~Munoz,
Phys.\ Lett.\ B {\bf 389}, 305 (1996) ;
S.~A.~Abel and C.~A.~Savoy,
Phys.\ Lett.\ B {\bf 444}, 119 (1998) ;
S.~Abel and T.~Falk,
Phys.\ Lett.\ B {\bf 444}, 427 (1998).
\bibitem{Abel:2000bj}
S.~A.~Abel, B.~C.~Allanach, F.~Quevedo, L.~Ibanez and M.~Klein,
JHEP{\bf 0012}, 026 (2000).
\bibitem{Ibanez:2000bn}
L.~E.~Ibanez,
Class.\ Quant.\ Grav.\ {\bf 17}, 1117 (2000), and references therein.
\bibitem{scale} E.~Witten,
Nucl.\ Phys.\ B {\bf 471}, 135 (1996) ;
J.~D.~Lykken,
Phys.\ Rev.\ D {\bf 54}, 3693 (1996) ;
N.~Arkani-Hamed, S.~Dimopoulos and G.~Dvali,
Phys.\ Lett.\ B {\bf 429}, 263 (1998) ;
I.~Antoniadis, N.~Arkani-Hamed, S.~Dimopoulos and G.~Dvali,
Phys.\ Lett.\ B {\bf 436}, 257 (1998).
\bibitem{Polchinski:1995mt}
J.~Polchinski,
   Phys.\ Rev.\ Lett.\ {\bf 75}, 4724 (1995).
\bibitem{Gabrielli:2001uy}
E.~Gabrielli, S.~Khalil, C.~Munoz and E.~Torrente-Lujan,
  Phys.\ Rev.\ D {\bf 63}, 025008 (2001).
\bibitem{cerdano}
See also 
D.~G.~Cerdeno, E.~Gabrielli, S.~Khalil, C.~Munoz and E.~Torrente-Lujan,
  hep-ph/0102270.
\bibitem{drees}
A.~Djouadi, M.~Drees and J.~L.~Kneur,
hep-ph/0107316.
\bibitem{ch} G.~Abbiendi {\it et al.} [OPAL Collaboration], 
Eur. Phys. J. {\bf C8} 255 (1999)
\bibitem{glue} P. Abreu {\it et al.} [DELPHI Collaboration], 
Phys. Lett. B {\bf 144} 491 (1998).
\bibitem{st1} S. Navas-Concha, talk given at SUSY 98, Keble College, 
Oxford, UK, July 1998. 
\bibitem{stau1} LEP SUSY Working Group, ALEPH, DELPHI, L3 and OPAL 
experiments, note LEPSUSYWG/98-02.1 . 
\bibitem{Cleo}
S.~Ahmed, {\it et al.} [CLEO Collaboration], CLEO-CONF-99-10, hep-ex/9908022.
\bibitem{bsg}
K.~Chetyrkin, M.~Misiak and M.~Munz,
Phys.\ Lett.\ B {\bf 400}, 206 (1997) ; (E) B {\bf 425}, 414 (1998) ;
A.~Ali and C.~Greub,
Phys.\ Lett.\ B {\bf 361}, 146 (1995) ;
C.~Greub, T.~Hurth and D.~Wyler,
Phys.\ Rev.\ D {\bf 54}, 3350 (1996) ;
N.~Pott,
Phys.\ Rev.\ D {\bf 54}, 938 (1996) ;
A.~L.~Kagan and M.~Neubert,
Phys.\ Rev.\ D {\bf 58}, 094012 (1998) ;
Eur.\ Phys.\ J.\ C {\bf 7}, 5 (1999).
\bibitem{future} S. Baek and P. Ko, in preparation.
\bibitem{largetan} G.~Degrassi, P.~Gambino and G.~F.~Giudice,
JHEP{\bf 0012}, 009 (2000) ;
M.~Carena, D.~Garcia, U.~Nierste and C.~E.~Wagner,
Phys.\ Lett.\ B {\bf 499}, 141 (2001).
\bibitem{dm}
M.~Drees and M.~M.~Nojiri,
Phys.\ Rev.\ D {\bf 47}, 4226 (1993) ;
M.~Drees and M.~Nojiri,
Phys.\ Rev.\ D {\bf 48}, 3483 (1993) ;
G.~Jungman, M.~Kamionkowski and K.~Griest,
Phys.\ Rept.\ {\bf 267}, 195 (1996) ;
A.~Djouadi and M.~Drees,
Phys.\ Lett.\ B {\bf 484}, 183 (2000) ;
M.~Drees, Y.~G.~Kim, T.~Kobayashi and M.~M.~Nojiri,
hep-ph/0011359 ;
S.~Y.~Choi, Seong-Chan~Park, J.~H.~Jang and H.~S.~Song,
hep-ph/0012370 ;
S.~Y.~Choi,
hep-ph/9908397.
\bibitem{dmexp}
R.~Bernabei {\it et al.} [DAMA Collaboration], Phys.\ Lett.\ B 
{\bf 480}, 23 (2000) ;
R.~Abusaidi {\it et al.} [CDMS Collaboration], Phys.\ Rev.\ Lett.\ 
{\bf 84}, 5699 (2000);
For more analysis, see A.~Bottino, F.~Donato, N.~Fornengo, S.~Scopel, 
Phys.\ Rev.\ D {\bf 62} 056006 (2000).
\bibitem{cdms}
R.~Schnee, talk at COSMO 2000, Cheju, 
Korea, Sep. 2000.
\bibitem{cresst}
M.~Bravin {\it et al.} [CRESST Collaboration], Astropart.\ Phys.\ 
{\bf 12}, 107 (1999).

\end{thebibliography}
\end{document}